\documentstyle[aps,pre,epsf,multicol]{revtex}
\begin{document}
\draft
\title{Two Dimensional Equilibrium Surface Roughness
for Dissociative Dimer Dynamics}
\author{Deok-Sun Lee$^{1,2}$ and Marcel den Nijs$^1$}
\address{
$^1$Department of Physics, University of Washington, Seattle,
Washington 98195, USA\\
$^2$School of Physics, Seoul National University, Seoul 151-747, Korea}
\maketitle
\begin{abstract}
Equilibrium crystal surfaces, constrained to equilibrate 
by means of dissociative dimer deposition and evaporation,
have anomalous global surface roughness.
We generalize earlier results for one dimensional interfaces
to two dimensions.
The global surface width scales with surface size $L$
as $ W^2\sim\log [L/ (\log L)^{\frac{1}{4}}]$
instead of the conventional form $W^2\sim\log L$.
The surface roughening  transition does not change in nature, 
but its location is subject to a large and slowly varying 
logarithmic finite-size-scaling shift.
\end{abstract}

\begin{multicols}{2}
\section{introduction}
Every now and then a new direction of research
sheds an unexpected light on an older topic.
In this instance, that older topic is equilibrium surface roughness,
and a study of non-equilibrium driven surface growth triggered it.
One dimensional (1D) surfaces lack equilibrium phase transitions,
but while being driven their stationary growing states can undergo
roughening transitions.
A recent example is directed-percolation-type roughening~\cite{alon}.
While generalizing the latter to directed-Ising-type
roughening~\cite{noh00,hinr},
Noh, Park, and den~Nijs discovered that the equilibrium point 
in their phase diagram had unusual properties.
The surface width $W$,
\begin{equation}
W^2(L) = \frac{1}{L} 
\sum_{\bf r} \left(\langle h_{\bf r}^2\rangle 
- \langle h_{\bf r}
\rangle^2 \right)
\simeq L^{2 \alpha},
\label{width}
\end{equation}
did not scale  with the conventional (random walk)
exponent $\alpha=\frac{1}{2}$,
but instead with what seemed to be a value close to
$\alpha=\frac{1}{3}$.
The origin of this is a global conservation law, 
an evenness constraint on the occurrences of every surface height, 
due to the dissociative dimer deposition and evaporation dynamics.
They related this to even-visiting random walks (RW)~\cite{cicuta},
i.e., RW required to visit 
every site an even number of times.
In follow-up studies~\cite{noh01,bauer}, it  was shown that
the value $\alpha=\frac{1}{3}$ is exact,
using a mapping to Lifshitz tails~\cite{lutt} in the density
of states of  1D fermions in  non-Hermitian random fields.

In this paper we address the issue whether something similar happens
in  two dimensional (2D) surfaces.
We show that the global surface roughness is again anomalous, 
i.e., that instead of the conventional
logarithmic finite-size divergence, $W^2\sim\log L $, 
the width diverges
as $W^2\sim \log L_{\rm free}$ with an effective surface size
$L_{\rm free}\sim L/(\log L)^\frac{1}{4}$. 
Moreover, the surface roughening
transition temperature has a large and slowly converging logarithmic 
finite-size-scaling correction.

The outline of this paper is as follows.
In Sec.~\ref{model} we review the most basic version of dissociative 
dimer dynamics
leading to the global evenness constraint.
In Sec.~\ref{2xL} we generalize the 1D results of Noh {\it et~al.} to 
semi-infinite
$N\times L$  lattices and show that the width still scales with 
$\alpha=\frac{1}{3}$.
This involves a representation of the equilibrium surface 
as a RW inside a $N$ dimensional tube.
In Sec.~\ref{LxL} 
we present our numerical data for a $L\times L$ lattice and
generalize the so-called healing time argument of Noh {\it et~al.} 
from 1D to 2D.
Next, in Sec.~\ref{randomfield},
we map the 2D evenness-constrained  equilibrium surface 
partition function onto that of a 2D surface with annealed
random fields which couple to the surface heights, 
and use this to derive
the relation  $L_{\rm free}\sim L/(\log L)^\frac{1}{4}$ analytically.
In Sec.~\ref{roughening} we discuss the implications of the anomalous 
roughness to the
roughening transition, and in  Sec.~\ref{exp} we comment on possible 
experimental realizations.
We summarize our finding in Sec.~\ref{summary}.

\section{dissociative dimer dynamics}\label{model}

The dynamic process that gives rise to the global evenness constraint
has in its bare-bone version the following  structure.
Consider the so-called restricted solid-on-solid (RSOS) model
on a 1D or 2D lattice. Every lattice site contains an integer-valued
height variable $h_{\bf r}=0,\pm1,\pm2,\cdots$, and nearest neighbors
can differ only by $\Delta h=0, \pm1$. The latter means that only 
surface steps of
height-one are allowed. The RSOS model has a long history in
the theory of surface roughening transitions 
(see e.g., ref.~\cite{mdn85}).

Impose now the following dimer-type deposition and evaporation
rule on the RSOS model.
Particles can arrive on and leave the surface 
only in pairs (as dimers).
Choose at random two nearest neighbor sites on the lattice.
The surface heights of both sites are increased or decreased
by one unit (with equal probabilities)
but only when those two sites are at equal heights, and
when the new configuration would not violate the RSOS rule
(only single height steps).
This simulates the deposition and the evaporation of
horizontal dimers.
The dimers are dissociative in the sense that the identity
of a dimer is lost after it is deposited, and that a particle
can arrive on and leave the surface with a different partner.

This type of dynamics was studied recently 
in one dimension~\cite{noh00}
and it was found that the equilibrium surface width
does not scale with  $\alpha=\frac{1}{2}$,
as in conventional  surface  dynamics,
but with the anomalous exponent $\alpha= \frac{1}{3}$
instead.
The value $\alpha=\frac{1}{2}$ is highly generic.
Equilibrium long range order cannot exist in 1D,
and therefore the probability to make an up or down step
while walking along an equilibrium surface is uncorrelated
beyond a definite correlation length.
This implies that the interface width scales
like the dispersion of a random walker, 
i.e., that  $\alpha=\frac{1}{2}$.

Dissociative  dimer dynamics circumvents this generic picture 
by imposing a non-local constraint on the Gibbs distribution. 
Particles are
deposited at each height level in pairs, and therefore the
RW must visit each level an even number of times. This is a
non-local constraint, due to the  dissociative character of the
dynamics. For example, on a  locally  flat surface segment, two
dimers are able to land next to each other, but next, the two
particles in the middle are allowed to switch partners and
evaporate together leaving two disconnected monomers behind.
Repeated processes like this leave no local trace of the dimer
nature of the dynamics, but globally every
height level needs to be occupied an even number of times. In
contrast,  non-dissociative dimer dynamics obeys the constraint
locally and therefore yields  the conventional scaling exponent
$\alpha=\frac{1}{2}$, just like monomer dynamics.

\section{semi-infinite lattices}\label{2xL}

We start with a generalization of the one dimensional $1\times L$ 
lattice results
of Noh {\it et~al.} to $N\times L$ lattices with $N=2,3,\cdots$.
On the one hand, 
having several channels instead of one weakens the evenness
constraint and maybe sufficiently 
to alter the $\alpha=\frac{1}{3}$ exponent.
On the other hand,  critical exponents change typically only with
dimension, and these $N$ channel lattices are still 1D systems
at large length scales.

\begin{figure}
\centerline{\epsfxsize=9cm \epsfbox{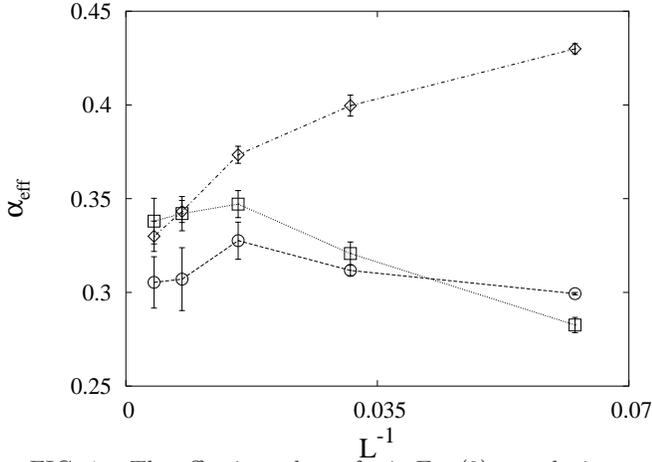}}
\caption{
The effective values of $\alpha$ in Eq.~(\ref{a_eff}) v.s. the inverse
of $L$  are shown for $1\times L$ surfaces (diamonds),  
$2\times L$ surfaces (circles), 
and $3\times L$ surfaces (squares) with  
$L=16$, $32$, $64$, $128$, and $256$.
}
\label{alpha_NxL}
\end{figure}

Our simulation results are shown in Fig.~\ref{alpha_NxL}.
We plot the effective values of $\alpha$
as functions of the system size $L$ for $N=2$ and $3$, evaluated as
\begin{equation}
\alpha_{\rm eff}(L)=\frac{\log\left[W(L)/W(\frac{1}{2}L)\right]}{\log 2}.
\label{a_eff}
\end{equation}
The roughness exponent remains close to $\alpha =\frac{1}{3}$
for both  $N=2$ and $N=3$.
The convergence is not impressive but 
comparable to that in the original
one channel case. These large finite-size-scaling corrections
are explained at the end of this section.

The numerical results for $\alpha$ suggest we seek an analytical 
generalization of the even-visiting random walk representation of
the system and from that show that the surface roughness
exponent does indeed not change.
Consider a walk along the equilibrium $2\times L$  
surface from one edge to the other
(from $x=0$ to $x=L$). The $x$ coordinate plays the role of time
and going up or down along the surface is the RW aspect.
The walk does not obey the evenness constraint
along one specific channel, $y=1$ or $y=2$, 
because dimers can be deposited
in two orientations, along and perpendicular to the channel.
However, if you walk twice, one time in each channel,
then the constraint is still obeyed for the two combined walks.
The evenness constraint is weakened, but obviously still present.

On a formal level, the connection between the random walk and 
surface roughness follows from the well known transfer matrix 
formalism for evaluating the partition function  of the surface.
Let's first consider a 1D lattice with
monomer-type dynamics (no evenness constraint).
The labels $n$ and $n_0$ in the partition function  $Z(n|n_0)_x$
represent fixed boundary conditions of the
surface at both ends, $h=n_0$ at site $0$, and $h=n$ at site $x$.
$Z(n|n_0)_x$ obeys the recursion relation
\begin{equation}
Z(n|n_0)_{x+1} = Z(n|n_0)_x + Z(n-1|n_0)_x +Z(n+1|n_0)_x,
\label{1D-rw}
\end{equation}
which  is a discrete version of the diffusion equation and can be
reinterpreted as the time evolution of a  1D random walker 
with $Z(n|n_0)_x$ the unnormalized probability to find the walker
at position $n$ at time $t=x$ starting from position $n_0$ at time
$t=0$.

Let's generalize this to the two channel lattice.
$n$ now becomes a two-component vector, ${\bf n}=(n_1,n_2)$.
The transfer-matrix-type recursion relation
\begin{equation}
Z(n_1,n_2|{\bf n_0})_{x+1} = \sum_{m_1,m_2}
Z(n_1-m_1,n_2-m_2|{\bf n_0})_x,
\label{2D-rw}
\end{equation}
could be  easily solved numerically because the transfer matrix is 
still finite.
However, for the sake of the evenness constraint, we will pursue that
Eq.~(\ref{2D-rw}) resembles again a diffusion equation
and again can be reinterpreted as the time evolution of a RW;
but this requires some care and the introduction of an absorbing wall.

This is a walk on a 2D lattice $(n_1,n_2)$.
The dimension of the walker's space is equal to the number of
channels, not to the dimension of the surface.
$h(1,x)$ represents the $n_1$ coordinate and 
$h(2,x)$ the $n_2$ coordinate of the walker at time $t=x$.
For example, when the walker is located at site $(4,5)$ at time
$t=10$,  the surface height at site $(1,10)$ is
equal to  $h=4$, and at site $(2,10)$ equal to  $h=5$.

\begin{figure}
\centerline{\epsfxsize=7cm \epsfbox{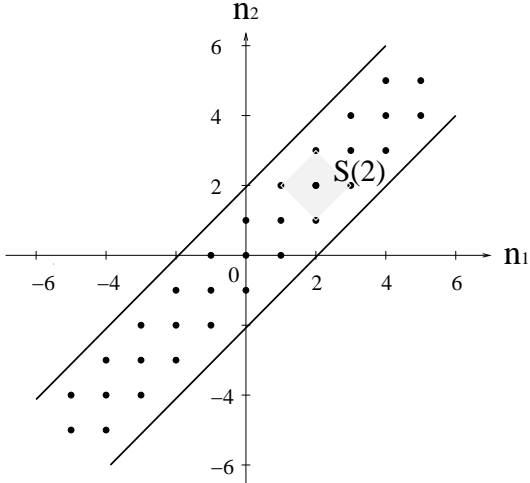}}
\caption{
The $2\times L$ (2 channel) equilibrium surface representation 
as a random walk on a strip with absorbing walls.
The surface coordinate along the channel represents
time in the random walk. The surface heights in the
first and second channel  represent the 
$n_1$ and $n_2$ coordinates of the walker.
The shaded area is an example of a plaquette with 
Ising spin $S(2)=\pm 1$.
The heavy solid lines are the absorbing walls.
See the text for more details.
}
\label{strip}
\end{figure}

The integer-valued summation labels $m_1$ and $m_2$ in 
Eq.~(\ref{2D-rw}) are subject to the RSOS condition that
nearest neighbor surface columns can differ only by $\Delta h=0,\pm1$.
This translates into two restrictions:
Firstly, at every  moment in time, the
two coordinates of the walker can differ only by $n_2-n_1=0,\pm1$.
So the walk is restricted in the $(n_1,n_2)$-plane to a strip 
centered along the diagonal, $n_1=n_2$,  
as illustrated in Fig.~\ref{strip}.
In other words, the walk remains quasi one dimensional after all.
Secondly, due to the same RSOS rule,
each positional component $n_1$ and $n_2$ of the random walk 
can change only by
$0$ or $\pm 1$ during each time step, $t=x\to x+1$.
It is easy to see that the walker can jump to only 6 new positions 
from any diagonal site, $n_1=n_2$, 
and to only 5 from every off-diagonal one, $n_1\neq n_2$
(in addition to not moving at all).

In the equilibrium Gibbs distribution 
for the RSOS monomer-type surface dynamics,
all surface  configurations have equal probabilities.
This is reflected in Eq.~(\ref{2D-rw}) by the fact that all transition
probabilities are equal.
However, this recursion relation does not represent yet a proper
Master equation for a random walk 
because it does not conserve probability.
A Master equation needs to be of the generic form:
\begin{equation}
Z({\bf n}|{\bf n_0})_{t+1}= \sum_{{\bf n}^\prime}
w({\bf n}^\prime,{\bf  n}) Z({\bf n}^\prime |{\bf n_0})_t,
\end{equation}
and the transition probabilities, $w({\bf n}^\prime,{\bf  n})$,
must conserve probability from any state $\bf n^\prime$, i.e.,
\begin{equation}
\sum_{\bf n} w({\bf n}^\prime,{\bf n}) =1.
\label{norm}
\end{equation}
The one channel transfer matrix equation of motion, 
Eq.~(\ref{1D-rw}) is of this form;  
except that we have to divide all transition
probabilities by a common factor, equal to $3$.
The two channel transfer matrix equation of motion, 
Eq.~(\ref{2D-rw}) does not satisfy  Eq.~(\ref{norm}).

Partition functions scale in `time' $t=x$ exponentially as
\begin{equation}
Z({\bf n}|{\bf n_0})_x \sim \lambda_0^x,
\end{equation}
with $\lambda_0$ the largest eigenvalue of their transfer matrices.
In Eq.~(\ref{1D-rw}),  $\lambda_0$ is simply equal to $3$, i.e., the
same common factor that normalizes  the transition probabilities.
Such a simple common factor does not exist for Eq.~(\ref{2D-rw}) 
because the numbers of sites to which the walker can hop are 
different at diagonal and off-diagonal sites.
This lack of conservation of probability
can still be incorporated in the random walk representation by 
interpreting the boundaries of the strips as so-called absorbing walls.
In the absence of the walls, the common factor would have been equal
to $3^2$. The remaining exponentially decaying factor 
$(\lambda_0/3^2)^{x}$ represents the absorption by the walls.
Moreover, $\lambda_0^x$-type factors drop out of the
calculation of thermodynamic averages, which in
the random walk representation is equivalent to
performing all averages with respect to the so-called
surviving ensemble only.

In a free 2D random walk,
the fluctuations in the two components decouple.
They do so also in our random walk with 
absorbing walls on the strip,
because we average over the surviving ensemble only.
This implies that in the direction parallel to the strip,
the root-mean-square displacement scales
with the conventional power as
\begin{equation}
\xi_\parallel (t)\sim t^{\frac{1}{2}},
\end{equation}
at time scales much larger than a characteristic  correlation time
arising from the range of hopping distances of the RW
during each time step. The root-mean-square
displacement in the perpendicular
direction, $\xi_\perp(t)$ remains finite due to the presence
of walls.
Thus we recover (although in a somewhat contrived manner)
the well known result that for a semi-infinite  $2\times L$ lattice
with $L\to\infty$, the equilibrium surface width still
scales as in 1D, with
$W= [(\xi_\parallel^2 (L) +\xi_\perp^2 (L))/2]^{\frac{1}{2}}
\sim L^{\frac{1}{2}}$.

Let's return now to the reason for constructing this random walk 
interpretation,
i.e., to deal with the evenness constraint.
For the one channel lattice,
the core  step towards the analytic derivation of $\alpha=\frac{1}{3}$
was the introduction of Ising variables to keep track of where the RW 
has been before.
An Ising spin $S_n=\pm 1$ was associated with each site, 
and flipped for each time interval the RW resided on that site.
The evenness constraint was enforced by
requiring that at time $t=x$ all spins point in the same direction as
at time $t=0$.
We need only to establish a generalized equivalent formulation
to demonstrate that the surface roughness exponent 
retains its 1D value
$\alpha=\frac{1}{3}$ on the two channel lattice.

Introduce an Ising spin $S(n)=\pm 1$ to each diagonal site $(n,n)$,
and associate it with the plaquette consisting of sites
$(n-1,n)$, $(n,n+1)$, $(n+1,n)$, $(n,n-1)$, and $(n,n)$ on the strip
(shown as the shaded area for $n=2$ in Fig.~\ref{strip}).
This spin is flipped for each unit time interval the walker resides on
$(n-1,n)$, $(n,n+1)$, $(n+1,n)$, or $(n,n-1)$.
$S(n)$ does not flip 
when the walker resides on the diagonal site $(n,n)$
because then the number of occurrences of height $n$ 
increases by an even number, $h(1,x)=h(2,x)=n$.
Note that two Ising spins flip
when the walker occupies an off-diagonal site, i.e., $S(n)$ and
$S(n+1)$ for the site $(n,n+1)$.

In the one channel case, the Ising spins act as gate keepers along
the chain, and the RW cannot pass site $n$ without
flipping its spin $S(n)$. Similarly, in the two channel case, the
RW moves on the quasi 1D strip and again a
line of Ising spins act as gate keepers. Topologically and at large
length scales, the situations are equivalent and therefore
$\alpha$ must be the same as in 1D.

One detail seems different.
On the chain, the walker is unable to pass site $n$
without flipping its spin $S(n)$, while on the strip,
it can pass through
plaquette $n$ without flipping $S(n)$. That might seem an
essential difference, but it is not.
A relaxed model where a spin is flipped stochastically
when the walker passes
was shown by Noh {\it et~al.} in the context of the purely 1D surface
to belong to the
same universality class as the deterministic spin flip model.
This followed most clearly from
the transfer-matrix equivalence to Lifshitz tails 
in the density of states
of 1D fermions in non-Hermitian random fields.
The random walker is the fermion and the Ising spins generate the
random potential (see ref.\cite{noh01} for details).

The generalization to $N\times L$ surfaces is straightforward.
The RW moves in a $N$ dimensional tube
(i.e., still remains effectively 1D) and the
evenness constraint still requires only a single line of Ising
spins positioned along the body-diagonal of the tube. Each flips
when the walker comes within range; $S(n)$ flips every
time any coordinate of the RW position is equal to $n$.
We conclude from the above evenness-constrained RW construction that
the roughness exponent must indeed retain its 1D value
$\alpha=\frac{1}{3}$ for all finite $N$.

Let's return to our numerical results.
The shapes of the curves in Fig.~\ref{alpha_NxL} represent
so-called finite-size-scaling corrections.
These corrections are  larger and more complex
than for the purely 1D one channel lattice~\cite{noh00}.
This can be understood qualitatively as follows.
For $N=1$, the evenness constraint is
ineffective at  small $L$
such that $\alpha_{\rm eff}$ (see Eq.~(\ref{a_eff})) 
decreases monotonically from a value  near the free
unconstrained  RW exponent $\alpha=\frac{1}{2}$ at small $L$ towards
$\alpha=\frac{1}{3}$ at $L\to \infty$.
(A characteristic crossover system size $L$ could
be constructed  in terms  of the  ratio between $L$
and length scale $L_{\rm free}(L) $ over which the surface
fluctuations are unimpeded by the evenness constraint;
$L_{\rm free}(L) $ is
defined in the healing time argument of Sec.~\ref{LxL}.)

In contrast, the $N=2$ and $N=3$ curves start off 
from a value near zero,
overshoot $\alpha=\frac{1}{3}$, and then bend backwards.
At small $L<N$, the surface behaves as if it is 2D
and as if the evenness constraint is absent,
i.e., $W\sim (\log L)^{\frac{1}{2}}$.
This explains why the curves start off at $\alpha\simeq 0$.
The surface starts to behave one dimensional beyond $L\sim N$,
but initially remains  still unaware of the evenness constraint,
such that the $\alpha$ curves in Fig.~\ref{alpha_NxL}
overshoot (towards $\alpha\simeq \frac{1}{2}$)
and only  back over at larger $L$ 
where the evenness constraint kicks in.

\section{Roughness on $L\times L$ lattices}\label{LxL}

We now turn our attention to dissociative dimer dynamics
on a truly 2D $L\times L$ lattice.
For monomer-type dynamics, it is well known that the height-height
correlation function diverges logarithmically, 
\begin{equation}
g({\bf r})=\langle ( h_{{\bf r}+{\bf r_0}}-h_{\bf r_0})^2\rangle
\simeq \frac{1}{\pi K_G} \log r,
\label{2D-rough}
\end{equation}
and the surface width scales as
\begin{equation}
W^2(L) \simeq \frac{1}{2\pi K_G} \log L,
\label{2D-width}
\end{equation}
with $K_G$ the so-called effective Gaussian coupling constant 
which in the RSOS model  at $T\to \infty$ (our numerical simulations)
takes the specific value $K_G\simeq 0.9$~\cite{mdn85}.
The goal is to find out how the evenness constraint affects this
on a global scale.

\begin{figure}
\centerline{\epsfxsize=9cm \epsfbox{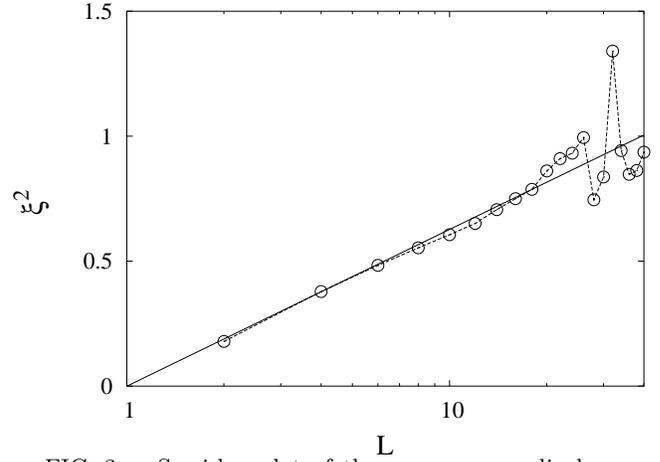}}
\caption{
Semi-log plot of the mean-square displacement $\xi^2(L)$ of the
random walk in the tube described in Sec.~\ref{LxL} as a function of
time $L$ for $2\leq L \leq 40$ with the total random walk time $L$
equal to the dimension of the tubular random walk space $N$.
The solid line fits the form $\xi^2(L) = 0.272\log L$
as predicted by Eq.~(\ref{log}).
}
\label{nu_LxL}
\end{figure}

In the random walk representation of the previous section, we need
to take the limit where the dimension of the tube ($N$) and
the random walk time ($L$)
diverge simultaneously, as $N=L\to \infty$.
This diminishes
the usefulness of the RW representation.
For example, according to Eq.~(\ref{2D-width}), the
root-mean-square displacement of the RW along the
body-diagonal of the tube must scale (for monomer dynamics) as
\begin{equation}
\xi_\parallel(L) \sim (L\log L)^{\frac{1}{2}}.
\label{log}
\end{equation}
This might seem at odds with the canonical $\sqrt L$ power
implied by the RW interpretation; but only until
one remembers that the walk is truly random only at time scales
$L$ much larger than a characteristic correlation time 
arising from the maximum hopping distance
of the walker during each time step, and realizes that
the latter (projected along the tube's body-diagonal)
is proportional to $\sqrt{N}$, and thus diverges
simultaneously with $L$.
For the sake of curiosity, we checked and confirmed
Eq.~(\ref{log}) numerically.
Fig.~\ref{nu_LxL} shows the average root-mean-square displacement,
$\xi(L)=[(\xi^2_\parallel(L)+\xi^2_\perp(L))/L]^{\frac{1}{2}}$ 
of the RW as a function of $L=N$.

Before we present our numerical results for the evenness-constrained
surface width in 2D, we like
to present an educated guess of what the behavior might be by
generalizing the so-called healing length argument
of Noh  {\it et~al.}~\cite{noh00}
from 1D to 2D. This argument was the least rigorous of their 
analytical results,
but explained the anomalous exponent $\alpha=\frac{1}{3}$ 
at a simple intuitive level.
In 1D and also for all our semi-infinite $N\times L$
surfaces of the previous section, the argument runs (in a somewhat
modified form) as follows.

Consider the semi-infinite $N\times L$ surface.
There must exist a crossover length  scale $L_{\rm free}$,
larger than the  surface width $N$
but much smaller than the total channel length $L$,
within which the evenness constraint can be ignored and
the surface fluctuates freely such that
the surface width increases as
\begin{equation}
W(L_{\rm free}) \sim {L_{\rm free}}^{\frac{1}{2}},
\label{xi1d-free}
\end{equation}
just like in 1D monomer dynamics.
The basic premise of the healing length
argument is that at larger length scales the surface  must be
preoccupied satisfying the evenness constraint such that the
surface width does not increase any further. Then, we need only 
to find out how $L_{\rm free}$ scales with $L$.

Divide the surface strip  $0<x<L$ into 
segments of length $L_{\rm free}$.
Choose one of these segments and mark all surface heights that
are visited an odd number of times in that segment.
It is known that these defect surface heights are 
uniformly  distributed
within the range $|h|< {\cal O}(W)$ such that their number is also
proportional to $W$~\cite{noh01}.
Suppose that the surface compensates for these defects one-by-one 
along the rest of the strip
(ignoring that the additional surface fluctuations introduce new
defects as well as repair simultaneously other old ones).
The average root-mean-square  distance the surface must
fluctuate vertically to reach that specific defect height 
is again proportional
to $W$, and that requires a typical horizontal length interval
$\Delta x\sim L_{\rm free}$.
So every segment  of length $L_{\rm free}$
leads to only one repair on average. Since there need to be  $W$ of
them, it follows that
\begin{equation}
\frac{L}{L_{\rm free}} \sim W,
\end{equation}
and, using Eq.~(\ref{xi1d-free}),  that
\begin{equation}
L \sim W ~ W^2, \quad {\rm and} \quad W\sim L^\frac{1}{3},
\end{equation}
such that for all $N\times L$ lattices
the roughness exponent is equal to $\alpha=\frac{1}{3}$.

For the  generalization of this argument to $L\times L$ lattices,
we divide the 2D lattice in blocks of size 
$L_{\rm free}\times L_{\rm free}$.
Within each block the surface fluctuates as if 
the global evenness constraint
does not exist, and the surface width scales as
$W^2\sim (2\pi K_G)^{-1} \log L_{\rm free}$.
Assume that again, like in 1D,
the oddly-visited surface heights are uniformly distributed
and therefore that their number is proportional to $W$.
The same repairing scheme as above predicts
again one repair on average per block, i.e., that
\begin{equation}
\left(\frac{L}{L_{\rm free}}\right)^2 \sim W,
\end{equation}
and therefore that
\begin{equation}
L \sim W^{\frac{1}{2}} e^{ 2 \pi K_G W^2}.
\label{scaling}
\end{equation}
In the limit of large $L$, we can invert Eq.~(\ref{scaling})  to
\begin{equation}
W^2(L)\sim  \frac{1}{2\pi K_G} \log\left[\frac{L}{(\log
L)^{\frac{1}{4}}}\right].
\label{dimer}
\end{equation}
The crucial features are the
logarithms and the $\frac{1}{4}$ power. The former originate from the
logarithmic divergence of unconstrained surface roughness in
2D and the latter  from the  scaling of the
number of blocks, $(L/L_{\rm free})^D$, with dimension $D$.

Compared to the conventional logarithmic divergence, 
Eq.~(\ref{2D-width}),
the lattice size $L$ is replaced by a logarithmically modified
effective length, $L_{\rm free} \sim L/ (\log L)^{\frac{1}{4}}$.
Logarithmic effects are notoriously difficult to confirm
numerically, except when you expect them.

\begin{figure}
\centerline{\epsfxsize=9cm \epsfbox{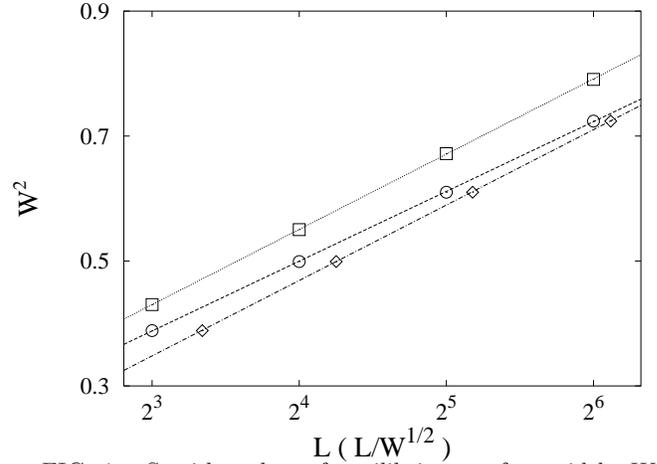}}
\caption{
Semi-log plots of 
equilibrium surface widths $W^2$ on $L\times L$ lattices for:
monomer dynamics (squares) and
dissociative dimer dynamics (circles)
both as functions of $L$.
The slopes of the straight-line fits yield
the values of $K_G$ (defined in Eq.~(\ref{2D-width})):
$K_G=0.916$ (monomer) and $0.988$ (dimer).
The diamonds show the same dimer dynamics data plotted as
$W^2$ v.s. $L/W^{\frac{1}{2}}$.
That line  has  nearly the same slope as the squares one 
(monomer dynamics), i.e., $K_G=0.914$ 
in agreement with Eq.~(\ref{dimer}).
}
\label{dimer_W}
\end{figure}

In Fig.~\ref{dimer_W}, we present our numerical simulation results.
The equilibrium surface width $W^2$ for monomer and dimer
dynamics  are plotted on  semi-log scales as functions of the system
size $L$ for $L=8$, $16$, $32$, and $64$ 
(squares and circles, respectively). 
The slope of the monomer curve  confirms the value
$K_G\simeq 0.9$ of Eq.~(\ref{2D-width}).
The dimer roughness line is also quite
linear, but with a reduced slope, which means that its surface
roughness could be fitted (in this range of length scales) by a
simple logarithm, but with an enhanced effective $K_G$.
However, when we plot $W^2$ versus $L/W^{\frac{1}{2}}$ as suggested by
Eqs.~(\ref{scaling}) and (\ref{dimer}), 
the slope of the (again straight) line is 
the same as in  the monomer case, i.e., we regain the
correct value of $K_G$. Therefore we conclude that for
dissociative dimer dynamics, Eqs.~(\ref{scaling}) and
Eq.~(\ref{dimer}) are indeed correct.
In the next section, we put
this result on a more rigorous analytic footing.

\section{surface roughness in random media}\label{randomfield}

Solid-on-solid models come in several variations.
The RSOS model we used above, where nearest neighbor columns can
differ only by $\Delta h=0,\pm 1$, is most convenient for numerical
simulations. The so-called discrete Gaussian
model~\cite{chui76,jose77}, where the $\Delta h$-restriction is
relaxed and replaced by a Gaussian interaction energy
\begin{equation}
A[h_{\bf r}] = \frac{E}{k_BT} = \frac{1}{2} K
\sum_{\langle {\bf r},{\bf r}^\prime \rangle} 
(h_{\bf r}-h_{{\bf r}^\prime})^2,
\label{DG}
\end{equation}
between nearest neighbor columns, 
$h_{\bf r}$ and $h_{{\bf r}^\prime}$
(the meaning of the $\left< ~~\right>$
brackets in the summation), is more appealing in analytical 
discussions. In this section we use the latter.

The global evenness constraint can
be incorporated  in the partition function in the following
manner. Define integer-valued variables $v_h$, one for every
surface height. Notice that they are not associated with any specific 
lattice site.
Each $v_h$ is equal to the number of times surface height level $h$ 
appears in the specific configuration.
The partition function, subject to the global evenness constraint,
can be written as
\begin{equation}
{\cal Z} = \sum_{\{h_{\bf r}\}} \left[\prod_h
\frac{1}{2}(1+z^{v_h})\right]~\exp\left(-A[h_{\bf r}] \right),
\label{pf1}
\end{equation}
with fugacity $z=-1$.
Every configuration in which any of the $v_h$'s is odd 
receives zero weight.

Next, we introduce randomness degrees of freedom $c_h=0,1$ 
associated with
those same surface height levels and write the partition function as
\begin{equation}
{\cal Z} = \sum_{\{c,h\}} \exp \left[-\sum_h ( \mu c_h v_h+\log 2 )
-A[h_{\bf r}]\right],
\label{pf2}
\end{equation}
with $z=\exp(-\mu)$.
Eq.~(\ref{pf2}) can be interpreted as the partition function of
an equilibrium surface in
the presence of annealed random external fields $\{c_h\}$ that
suppress the occurrences of specific surface heights. The noise
is global in the sense that the random variable $c_h$
affects surface  height level $h$
equally at every position $\bf r$ along the surface.
For every occurrence of height level $h$, the Boltzmann weight is
multiplied by a factor $z$ when $c_h=1$ (and not if $c_h=0$).
The fugacity parameter
$z$ allows us to interpolate between monomer and dimer dynamics
because at $z=1$ we retrieve the conventional discrete Gaussian
partition function.

We are following in this closely the analogous formulation  
for the 1D surface by Noh {\it et~al.}~\cite{noh01}.  
They found that  in 1D
$z=0$ acts as a special point, a stable fixed point in the
sense of renormalization transformations with all $-1\leq z<1$ as
its basin of attraction. We will show that this remains
true in 2D.

The limit $z=0$ is special. The fugacity Boltzmann  factors become
delta functions,
$z^{c_h v_h}\to \delta(c_h v_h)$, such that  the
surface is strictly  prohibited to pass through all height levels for 
which $c_h=1$.
Each of them acts as an impenetrable barrier
and the surface is restricted in its vertical height fluctuations
to wander between two such randomly placed walls.
The partition function factorizes and reduces to
\begin{equation}
{\cal Z}= \sum_{\xi} P(2\xi) {\cal Z}_{\xi},
\label{dec}
\end{equation}
where $P(2\xi)$ is the probability of finding two neighboring $c_h=1$
walls at a distance $\Delta h= 2\xi$ apart. Walls are placed at random
with probability $\frac{1}{2}$, such that $P(2\xi)= 2^{-2\xi}$. ${\cal
Z}_{\xi}$ in Eq.~(\ref{dec}) is the partition function of a
surface  restricted to fluctuate between two such hard
walls:
\begin{equation}
{\cal Z}_{\xi}=\sum_{ \{ h_{\bf{r}} \} }
\left[\prod_{\bf{r}} \theta(\xi-|h_{ \bf{r} }| )
\right]~\exp(-A [ h_{\bf{r}} ] ).
\end{equation}
The two following modifications cannot affect
the large length scale behavior.
Firstly, below the equilibrium
roughening transition the integer-valued height variables
$h_{\bf r}=0,\pm 1,\pm 2,\cdots$,
can be approximated by real-valued continuous variables
with the coupling constant $K$ renormalized to its
Gaussian value $K_G$. This is well known from the theory of
conventional surface 
roughening~\cite{chui76,jose77,knops,rough-revs,mdn-king}.
Secondly, 
we can replace the hard impenetrable walls by soft ones 
and approximate ${\cal Z}_{\xi}$ by 
\begin{equation}
\widetilde{{\cal Z}}_{\xi}=\sum_{\{h_{\bf{r}}\}} 
\delta(\langle
h^2 \rangle -\xi^2) \exp(-A[h_{\bf{r}}]),
\end{equation}
where the trace runs over
ordinary configurations of the Gaussian model 
but restricted to the
subset where the root-mean-square
surface width in each configuration is equal to $\xi$.
The evaluation of $\widetilde{\cal Z}_{\xi}$ is elementary.
Compared to the Gaussian partition function,
${\cal Z}_0 = \sum_{\{h_{\bf{r}}\}} \exp (-A[h_{\bf{r}}])$,
it gains an exponential factor, 
\begin{eqnarray}
\widetilde{\cal Z}_{\xi} &\sim& \int d m \sum_{\{h_{\bf{r}}\}}
\exp\left[
-\frac{1}{2} K_G \sum_{\langle {\bf{r}},{\bf{r}}^{'}
\rangle}(h_{\bf{r}}-h_{{\bf{r}}^\prime})^2  \right.\nonumber \\
& &  \left. + 2\pi i m \sum_{\bf{r}}
(h_{\bf{r}}^2-\xi^2)\right] \nonumber \\ 
&\sim& {\cal Z}_0 \exp\left[
-a_0
L^2 \exp(-4\pi K_G \xi^2)\right],
\label{zapprox}
\end{eqnarray}
where $a_0=\frac{\pi}{16}$.
Finally, the surface width of the annealed ensemble average
follows from Eqs.~(\ref{dec}) and (\ref{zapprox}) using the method
of  steepest descent as 
\begin{eqnarray}
W^2 &\sim& \frac{\sum_{\xi} \xi^2
P(2\xi){\cal Z}_{\xi}}{\sum_{\xi}P(2\xi){\cal Z}_{\xi}} \nonumber
\\ &\sim& \frac{\sum_{\xi} \xi^2 \exp\left[-2\xi \log2 -a_0 L^2 \exp(-4\pi K_G
\xi^2)\right]}{\sum_{\xi} \exp\left[-2\xi \log2 -a_0 L^2 \exp(-4\pi K_G
\xi^2)\right]} \nonumber \\ &\sim& \frac{1}{2\pi K_G} \log\left[\frac{L}{(\log
L)^{\frac{1}{4}}}\right].
\label{dimer2}
\end{eqnarray}
This is the same scaling form as proposed in the
previous section from the healing argument, Eq.~(\ref{dimer}).

The above derivation remains equally valid for small values of $z$ 
around $z=0$. There, the  surface can tunnel through the walls, but
with exponentially small probabilities (proportional to the length
of the intersection contours) actually resembling  more closely the
soft walls in the above derivation.
At the positive $z$ side, it is intuitively reasonable
that $z=1$ (the monomer dynamics point) is the horizon and
limiting point for this type of scaling.
At the negative $z$ side, the horizon extends to and
apparently includes the point $z=-1$. The numerical results and
also the healing argument of the previous section suggest this.
This is also intuitively consistent.
Negative values of $z$ imply an imaginary chemical potential
and therefore a mix of positive and negative Boltzmann weights that cancel
out against each other; this enhances the exponentially decaying
nature of the $|z|^{v_hc_h}$ weight factors,
and preserves this decay  even at $z=-1$.

\section{Roughening Transitions}\label{roughening}

The surface is presumed to be above 
the roughening transition temperature
$T_R$ in most of the above discussion.
For example, although Eqs.~(\ref{pf1}) and (\ref{pf2}) apply to 
all $T$,
it is correct to approximate the discrete Gaussian height
variables by  continuous Gaussian ones only for $T>T_R$.
Everywhere in the rough phase the global surface roughness has a
logarithmic correction, Eq.~(\ref{dimer}),
to its conventional  simple logarithm.
The natural question arises whether and/or how this influences
the nature of the surface roughening phase transition.
In this section we will show that the evenness constraint has no effect
on the properties of the roughening transition 
in the thermodynamic limit,
but gives rise to strong finite-size-scaling corrections,
including an apparent shift in the transition temperature.

Equilibrium surface roughness has been a topic of active research
over several decades. The roughening is an example
of a Kosterlitz-Thouless (KT) transition~\cite{KT}.
Experimental realizations include
Helium crystal surfaces ~\cite{He-revs},
metal surfaces~\cite{metal-revs},
and organic crystals~\cite{org-revs}.

The anomalous roughness exists only at a global scale.
Within the crossover length scale 
$L_{\rm free}\sim L/(\log L)^{\frac{1}{4}}$
(defined in Sec.~\ref{LxL}),
the  surface fluctuates freely
in disregard of the global evenness constraint;
i.e.,  inside the bulk the equilibrium surface remains
indistinguishable from that in monomer dynamics,
and since phase transitions are ruled by the bulk,
the roughening transition must remain in the KT universality class.

The `evenness boundary effect' is, however, very strong.
We will approach this from the perspective of
the original argument by KT where they estimated the
transition temperature by the free energy  of a single
vortex in, e.g., superfluid films~\cite{KT}.
The discreteness of the height variables,
$h_{\bf r}=0,\pm1, \pm2, \cdots$, 
in the discrete Gaussian solid-on-solid model
plays the same role as those vortices.
This might be somewhat surprising, but follows
mathematically from a so-called duality transformation~\cite{knops}.
In the rough phase, the heights can be treated as continuous variables
and this represents the `vortex-free phase'.
Consider the so-called sine-Gordon  correlation function in the
Gaussian model, in which the $h_{\bf r}$'s are continuous,
\begin{eqnarray}
\langle \exp\left[2\pi i (h_{{\bf r}+{ \bf r_0}}-h_{\bf r_0})\right] \rangle
&=& \exp\left[- \frac{1}{2} (2 \pi)^2 g(\bf r)\right] \nonumber \\
&\sim& r^{-2 \pi/K_G},
\label{charges}
\end{eqnarray}
(see also Eq.~(\ref{2D-rough})).
The logarithm of this is the free energy of placing
two topological objects, $\exp(\pm 2 \pi i h)$,  
at a distance $r$ apart.
They  favor integer values  of $h$.
That free energy scales logarithmically with distance $r$, and
a single one costs therefore an amount of free energy  
$f= (\pi/K_G) \log L$
or $f= (\pi/K_G- 2)\log L$ if we allow it to be at any position
in 2D space. 
The temperature $K_G=\frac{\pi}{2}$ where the latter is equal to zero
is the famous  KT estimate for the transition temperature.
(Recall that the coupling constant $K_G$ is measured in units
of $k_BT$ such that $K_G\sim T^{-1}$.)
This is how this works for monomer dynamics.

For dimer dynamics the two sine-Gordon `charges'
interact logarithmically, just like in Eq.~(\ref{charges}) but only
at distances smaller than $L_{\rm free}$.
Closer to the lattice size the interaction levels off such
that a single excitation now costs
\begin{equation}
f\simeq  \frac{\pi}{K_G} \log L_{\rm free} - 2 \log L.
\end{equation}
This yields,  by setting $f=0$ and using Eq.~(\ref{dimer}),
\begin{equation}
K_G \simeq \frac{1}{2} \pi 
\left[1-\frac{b_0 + \log(\log L)^{\frac{1}{4}}}{\log L}\right],
\end{equation}
with $b_0$ a constant.
This is a significant finite-size-scaling correction,
both in magnitude and in the slowness with which it scales
with $L$.
The transition appears to happen at a higher temperature.
For example, at $L=1000$ (larger than
most experimental  surface heterogeneity
lengths~\cite{metal-revs}), the shift is of order
\begin{equation}
\frac{\Delta T_R}{T_R} \simeq \frac{
\log(\log L)^{\frac{1}{4}}}{\log L}\simeq 7\%.
\label{deltaTR}
\end{equation}

\section{Experimental Realizations}\label{exp}

The scope of this paper is foremost theoretical,
i.e., to show how a seemingly benign
topological conservation law in the dynamics
strongly affects the equilibrium state.
Experimental realizations of the evenness constraint
should be possible to establish, but this requires careful
considerations and collaborations with experimentalists.
For example, molecular-type bonded molecules like
$N_2$ look promising, but unfortunately,
in solid $N_2$ the molecules do not dissociate.
As a starting point we include here a section
with a brief discussion of the most essential features.

Simple models like ours are meant and perfectly suited to discover
fundamental scaling laws associated with surface growth, but
they are rather simple minded when it comes to making
direct contact with actual experiments.
Those detailed theories of crystal growth have reached
a high level of sophistication in recent years~\cite{surfgrowth-revs}.
The evenness effect represents a topological feature
which will be preserved with increased realism.
After identifying a suitable experimental system,
the evenness constraint needs to be embedded into the appropriate
more detailed theoretical description.
However, at this stage, it suffices to focus on general aspects,
in particular those that possibly upset the conservation law.

The condition in our model that the $X_2$ molecules 
land only horizontally
is not a serious constraint, 
because the molecule  must dissociate as well.
The latter requires that the binding energy
between the two atoms be weaker or 
at best be of the same order of magnitude
as the bonding energies inside the solid.
In that case, both atoms strongly prefer to be close 
to the surface, and a vertically landed molecule 
will quickly decay.

A more serious issue is surface diffusion, 
which is omitted in our model.
The evenness constraint is preserved as long as a surface
wanders around on the same terrace, but
is lost when surface atoms jump across steps to lower 
or higher levels.
Diffusion across steps is reduced  by the presence of so-called
Schwoebel barriers~\cite{schwoebel}.
If those barriers are strong enough, the jumps will occur
infrequently enough that the evenness constraint remains
satisfied at length scales large compared to the heterogeneity length
scale of the experimental surface.
The latter is the length scale
at which defects like impurities pin and limit the surface
dynamics; this length  rarely exceeds $1000{\rm \AA}$.

Moreover, recall that the anomalous surface roughness
is stable with respect to varying the fugacity parameter $z$ in 
Eq.~(\ref{pf1}),
i.e., it remains present  
when the evenness constraint is not strictly
obeyed but only statistically. Therefore it might well be that
diffusion across steps preserves the anomalous roughness 
beyond the above estimate. 
This issue  needs further theoretical study.

Finally, the search for experimental
realizations does not need to be limited to dimer-type dynamics.
The anomalous global surface roughness exists and is the same
for $X_n$-type dissociative dynamics with any $n>1$.
This was found to be true earlier in 1D~\cite{noh01},
and follows also in 2D from generalizing our
analytic arguments.

\section{Summary}\label{summary}

In this paper we studied how equilibrium surface roughness
in two dimensional  surfaces is affected by a global constraint
that every surface height be present an even number of times
in every configuration. We presented numerical and analytic evidences.

In semi-infinite $N\times L$ surfaces, the evenness constraint is
weakened compared to the $1\times L$ surface,
but not enough to change the global width.
It still scales anomalously as $W\sim L^{\frac{1}{3}}$, 
just like in  purely
1D surfaces~\cite{noh00,noh01}.

In truly 2D $L\times L$ surfaces, the global
surface roughness  is also anomalous,
but the effect is weaker. The surface width scales
logarithmically as
\begin{equation}
W^2 (L)\sim \log L_{\rm free},
\end{equation}
similar to conventional monomer dynamics, but with a
logarithmically corrected  effective surface size
\begin{equation}
L_{\rm free}\sim \frac{L}{(\log L)^{\frac{1}{4}}}.
\end{equation}
The surface roughening  transition does not change in nature, 
but its location is subject to a large and slowly varying logarithmic 
finite-size-scaling shift, Eq.~(\ref{deltaTR}).

These results remain valid when the evenness constraint is only
statistically obeyed.
Moreover, the analytic derivation in Sec.~\ref{randomfield} involves
a mapping onto a surface model with annealed randomly placed
barriers, placed horizontally to the surface,
that inhibit the vertical surface fluctuations.
The surface roughness of those surfaces behaves the same as
the evenness-constrained ones.

Finally, 2D represents a critical dimension for ordinary
equilibrium surface roughness. This follows trivially by
evaluating the surface roughness in the Gaussian approximation,
$W\sim L^{\alpha}$, with $\alpha= (2-D)/2$ (logarithmic in $D=2$).
The surface width does not diverge in dimensions  $D>2$ 
and the surface remains asymptotically flat.
Still, the  healing length  and random field arguments 
can be generalized to higher dimensions:
$L_{\rm free}\sim L^{\frac{2D}{2+D}}$ and
$W\sim (L_{\rm free})^\alpha\sim L^{D\frac{2-D}{2+D}}$.
So for example, in $D=3$,  
monomer dynamics yields $W\sim L^{-\frac{1}{2}}$
and dissociative dimer dynamics $W\sim L^{-\frac{3}{5}}$.
The evenness constraint always flattens the surface
at a global scale.

\acknowledgements
This research is supported by the National Science Foundation 
under grant DMR-9985806, and by the Brain Korea 21 Project.

\end{multicols}

\end{document}